\documentclass[5p]{elsarticle}

\usepackage{geometry}
\usepackage{hyperref}
\usepackage{amsmath,amssymb,amstext,amsfonts,bbm,algorithm,algorithmicx,graphicx,xspace,nicefrac}
\usepackage{color,stmaryrd,enumerate,latexsym,bm,amsfonts,
  subfigure,wrapfig,verbatim,tabularx,textcomp}
\usepackage[small]{caption}
\usepackage{comment} 
\usepackage{epsfig} 
\usepackage{latexsym,nicefrac,bbm}
\usepackage{xspace}
\usepackage{color,fancybox,graphicx,url,subfigure}
\usepackage{enumitem}
\usepackage{booktabs}
\usepackage{commath}
\usepackage{mdframed}

\newcommand{\defeq}{\stackrel{\textup{def}}{=}}

\newtheorem{theorem}{Theorem}
\newtheorem{lemma}[theorem]{Lemma}

\newtheorem{proposition}[theorem]{Proposition}
\newtheorem{fact}[theorem]{Fact}

\DeclareMathOperator*{\pr}{\bf Pr}
\DeclareMathOperator*{\av}{\bf E}

\newcommand{\tr}[1]{\mathop{\mbox{Tr}}\left({#1}\right)}

\newcommand{\nfrac}[2]{\nicefrac{#1}{#2}}

\renewcommand{\norm}[1]{\ensuremath{\left\lVert #1 \right\rVert}}

\newcommand\rea{\mathbb R}

\newcommand\calN{\mathcal{N}}

\newenvironment{proofof}[1]{\begin{trivlist} \item {\bf Proof
#1:~~}}
  {\qed\end{trivlist}}

\newcommand{\eps}{\varepsilon}

\newcommand{\Id}[1]{\ensuremath{\mathbbm{I}}_{#1}}

\newlength{\algtopspace}
\newlength{\algpostcaptionspace}
\renewcommand{\defeq}{:=}

\begin{document}

\author[SS]{Sushant Sachdeva\corref{cor1}}
\ead{sachdeva@cs.yale.edu}
\address[SS]{Yale University, New Haven, CT,
  USA.}

\author[NV]{Nisheeth K. Vishnoi}
\ead{nisheeth.vishnoi@epfl.ch}
\address[NV]{\'{E}cole Polytechnique F\'{e}d\'{e}rale
  de Lausanne (EPFL), Lausanne, Switzerland.}

\cortext[cor1]{Corresponding author}


\title{The Mixing Time of the Dikin Walk in a Polytope \\
 --- A Simple Proof}  \date{\today}
\begin{abstract} 
  We study the mixing time of the Dikin walk in a polytope --- a
  random walk based on the log-barrier from the interior point method
  literature. This walk, and a close variant, were studied by
  Narayanan (2016) and Kannan-Narayanan (2012). Bounds on its mixing
  time are important for algorithms for sampling and optimization over
  polytopes. Here, we provide a simple proof of their result that this
  random walk mixes in time $O(mn)$ for an $n$-dimensional polytope
  described using $m$ inequalities.
  \end{abstract}
\begin{keyword}
{Polytopes, Sampling, Volume computation}, Random walks, Interior point methods.
\end{keyword}
\maketitle

\section{Introduction}
Sampling a point from the uniform distribution on a polytope
$K \subseteq \rea^n$ is an extensively-studied problem and is a
crucial ingredient in several computational tasks involving convex
bodies.  Towards this, typically, one sets up an ergodic and reversible random walk
inside $K$ whose stationary distribution is uniform over $K$. 
  The
mixing time of such a walk determines its efficacy, and, in turn,
depends on the isoperimetric constant of $K$ with respect to the
transition function of the walk.  Starting with the influential work
of Dyer \emph{et al.}~\cite{DyerFK91}, 
there has been a long line of work on faster and
faster algorithms for generating an approximately uniform point from a
convex body. Moreover, since convex bodies show up in a variety of
areas, there is a wide body of work connecting random walks and
isoperimetry in convex bodies to several areas in mathematics and
optimization.

One such important connection to the interior point method literature
was presented in the works of Kannan and Narayanan \cite{KannanN2012} and Narayanan~\cite{Narayanan2016} who proposed
the {\em Dikin walk} in a polytope.  
Roughly, the uniform version of the Dikin walk,
  considered by \cite{KannanN2012}, when at a point $x \in K,$
  computes the {\em Dikin ellipsoid} at $x,$ and moves to a random
  point in it after a suitable Metropolis filter.  The Metropolis
step ensures that the walk is ergodic and reversible.
  The Gaussian version of the Dikin walk, considered by
  \cite{Narayanan2016}, picks the new point from a Gaussian
  distribution centered at $x$ with its covariance given by the Dikin
  ellipsoid at $x,$ and applies a suitable Metropolis
  filter. 
 The Dikin ellipsoid at a point $x$ is the ellipsoid
  described by the Hessian of the log-barrier function at $x.$ It was
  introduced by Dikin in the first interior point method for linear
  programming~\cite{Dikin67}.

  Several virtues of the Dikin ellipsoid (see \cite{NesterovN94,
    Renegar01, Karmarkar84}) were used by \cite{KannanN2012,
    Narayanan2016} to prove that the mixing time of the Dikin walk is
  $O(mn)$ starting from a warm start, when $K$ is described by $m$
  inequality constraints. Recall that a distribution over $K$ is said
  to be a warm start if its density is bounded from above by a
  constant relative to the uniform distribution on $K$.  Roughly, the
  proof (for either walk) consists of two parts: (1) an isoperimetric
  inequality, proved by Lov\'{a}sz \cite{Lovasz99}, for convex bodies
  in terms of a distance introduced by Hilbert, and (2) a bound on the
  changes in the sampling distributions of the Dikin walk in terms of
  the Hilbert distance. The bound in (2) was the key technical
  contribution of~\cite{KannanN2012, Narayanan2016} towards
  establishing the mixing time of the Dikin walk.  We present a simple
  proof of this bound for the Gaussian Dikin walk implying that it
  mixes in time $O(mn).$ Our proof uses well-known facts about
  Gaussians, and concentration of Gaussian polynomials.

\subsection{Dikin walk on Polytopes}
Suppose $K \subseteq \mathbb{R}^n$ is a bounded polytope with a
non-empty interior, described by $m$ inequalities,
$a_{i}^{\top}x \ge b_{i},$ for $i \in [m].$ We use the notation
$x \in K$ to denote that $x$ is in the interior of $K.$ The
log-barrier function for $K$ at $x \in K$ is
$F(x) \defeq - \sum_{i \in [m]} \log (a_{i}^{\top}x - b_{i}).$ Let
$H(x)$ denote the Hessian of $F$ at $x,$ \emph{i.e.},
$H(x) \defeq \sum_{i\in [m]} \frac{1}{(a_{i}^\top x - b_{i})^2 }
a_{i}a_{i}^{\top}.$
For all $x \in K,$ $H(x)$ is a positive definite matrix, and defines
the \emph{local norm} at $x,$ denoted $\norm{\cdot}_{x},$ as
$\norm{v}^{2}_{x} \defeq v^{\top} H(x) v.$ The ellipsoid
$\{z : \norm{z-x}_{x} \le 1\}$ is known as the Dikin ellipsoid at $x.$

From a point $x \in K,$ the next point $z$ in the Dikin walk is
sampled from the Dikin ellipsoid at $x.$ The uniform
  Dikin walk, considered by \cite{KannanN2012},
  sampled the new point $z$ from the {uniform
    distribution in this ellipsoid.}
In the Gaussian Dikin walk, considered by \cite{Narayanan2016},  $z$ is sampled
from $g_{x},$ a multivariate Gaussian distribution centered at $x$
with covariance matrix $\frac{r^{2}}{n} H(x)^{-1},$ where $r$ is a
constant.
Thus, the density of the distribution is given by
\[g_x(z) = {\sqrt{\det H(x)}} \left(\frac{n}{2\pi r^2}
\right)^{\nfrac{n}{2}}\cdot \exp\left( -\frac{n}{2r^2}
  \cdot\norm{z-x}_{x}^2\right).\]
Equivalently, the next point $z$ is given by
\[z = x + \frac{r}{\sqrt{n}} \left(H(x) \right)^{-\nfrac{1}{2}} g,\]
where $g$ is an $n$-dimensional vector with each coordinate of $g$
sampled as an independent standard gaussian $\calN(0,1)$.

{ In order to convert this into a random walk that stays inside $K,$
with its stationary distribution as the uniform distribution on $K$,}
we apply the Metropolis filter to obtain the transition probability
density $p_x$ of the {Gaussian} Dikin walk: $\forall z \neq x,$ if $z \in K,$
$p_x(z) = \min\{g_x(z),g_z(x)\}$ (the walk stays at $x$ with the
remaining probability).

\subsection{Hilbert Metric, Isoperimetry, and Mixing Time}
We introduce the distance function which plays an important role in establishing the mixing time of the Dikin walk. Given two points
$x,y \in K,$ let $p,q$ be the end points of the chord in $K$ passing
through $x,y,$ such that the points lie in the order $p,x,y,q.$ We
define
$\sigma(x,y) \defeq \frac{|\overline{xy}||\overline{pq}|}
{|\overline{px}||\overline{qy}|},$
where $|\overline{xy}|$ denotes the length of the line segment $xy.$
$\log (1+\sigma(x,y))$ is a metric on $K,$ known as Hilbert metric.

{ Lov\'asz proved the following theorem for any random walk on $K$:}
Suppose for any two initial points $x,y \in K$ that are close in
$\sigma$ distance, the statistical distance of the distributions after
one step of the walk each from $x$ and $y,$ is bounded away from
1. Then, the lazy version of the random walk (where we stay at the
current point with probability $\nfrac{1}{2}$ at each step) mixes
rapidly.
\begin{theorem}[Lov\'{a}sz~\cite{Lovasz99}]\label{thm:lovasz}
  Consider a reversible random walk in $K$ with its stationary
  distribution being uniform on $K.$ Suppose $\exists \Delta > 0$ such
  that for all $x,y \in K$ with $\sigma(x,y) \le \Delta,$ we have
  $\norm{p_{x} - p_{y}}_{1} \le 1 - \Omega(1),$ where $p_{x}$ denotes
  the distribution after one step of the random walk from $x.$ Then,
  after
  $O(\Delta^{-2} )$ steps, the lazy version
  of the walk from a warm start is within $\nfrac{1}{4}$ total
  variation distance from the uniform distribution on $K.$
\end{theorem}
\noindent
Kannan and Narayanan proved that the transition function of the
{uniform} Dikin walk, $p_x,$ for $x \in \mathrm{int}(K),$ satisfies the
hypothesis of the theorem above with $\Delta = \Omega\left(\frac{1}{\sqrt{mn}} \right)$, thus implying that it mixes in
$O(mn)$ steps from a warm start. 
An analogous result for the Gaussian Dikin walk is implicit in the work of Narayanan.
Our main contribution is an alternative and simple proof of their main technical contributions.
In particular, we prove  the following theorem.  
\begin{theorem}
\label{thm:main}
  Let $\eps \in (0,\nfrac{1}{2}].$ For the Gaussian Dikin walk on $K$ with
  $r \le \frac{\eps}{400} (\log \frac{200}{\eps} )^{-\nfrac{3}{2}},$
  for any two points $x, y \in K$ such that
  $\norm{x-y}_{x} \le \frac{r}{\sqrt{n}},$ we have
  $\norm{p_{x} - p_{y}}_1 \le \eps.$
\end{theorem}
In order to use this theorem along with Theorem \ref{thm:lovasz} to
obtain the claimed mixing time bound, one needs a simple fact that,
for any $x,y$ in a polytope $K,$ which is described using $m$
inequalities, $\sigma(x,y) \ge \frac{1}{\sqrt{m}} \norm{x-y}_{x}.$ A
proof of this fact is given in the appendix; see Lemma
\ref{lem:hilbert-local-metric}.

The following two lemmas are the main ingredients in the proof of
Theorem~\ref{thm:main}: (1) If two points $x,y$ are close in the local
norm, \emph{i.e}, $\norm{x-y}_{x} \le \frac{r}{\sqrt{n}},$
then the two Gaussian distributions $g_{x}$ and $g_{y}$ are close in
statistical distance. (2) If $r$ is small enough (as a function of
$\eps$), then for all $x,$ $p_{x}$ and $g_{x}$ are $\eps$-close in
statistical distance.
\begin{lemma}
\label{lem:gaussian-stat-dist}
Let $r \le 1,$ and $c \ge 0$ be such that $c \le \min\{r, \nfrac{1}{3} \}.$ Let
$x,y \in K.$ If $\norm{x-y}_{x} \le \frac{c}{\sqrt{n}},$ then
$\norm{g_{x}-g_{y}}_{1} \le 3c.$
\end{lemma}
\noindent
This lemma relies on a well-known fact about the Kullback-Leibler
divergence between two multivariate Gaussian distributions, and
Pinsker's inequality that bounds the statistical distance between two
distributions in terms of their divergence.
\begin{lemma}
\label{lem:metropolis-stat-dist}
Given $\eps \in [0,\nfrac{1}{2}],$ for
$r \le \frac{\eps}{100} (\log \frac{50}{\eps} )^{-\nfrac{3}{2}},$ we
have $\norm{p_x - g_{x}}_{1} \le \eps.$
\end{lemma}
\noindent
This lemma, which shows that the Metropolis filter does not
change the distribution much, relies on a result on the concentration
of Gaussian polynomials, proved using hypercontractivity.  Given the
above lemmas, Theorem~\ref{thm:main} follows by applying triangle
inequality.

\section{Statistical distance between Gaussians and the local norm}
In this section, we present a proof of
Lemma~\ref{lem:gaussian-stat-dist} that bounds the statistical
distance between $g_{x}$ and $g_{y}$ for two points $x,y$ that are
close in the local norm. We need the following well-known fact about
the Kullback-Leibler divergence between two multivariate Gaussian
distributions.

\begin{fact}
\label{fact:multigaussian-KL}
Let $G_{1} = \calN(\mu_{1},\Sigma_{1})$ and $G_{2} =
\calN(\mu_{2},\Sigma_{2})$ be two $n$-dimensional Gaussian
distributions. 
Then,
\begin{align*}
\textrm{D}_{\textrm{KL}}(G_{2} | | G_{1}) = \frac{1}{2} \left(
  \tr{\Sigma_{1}^{-1} \Sigma_{2}} -n + \log \frac{\det \Sigma_{1}}{\det
    \Sigma_{2}} \right.  \\ \left. + (\mu_{1} - \mu_{2})^{\top}
\Sigma_{1}^{-1} (\mu_{1} - \mu_{2}) \right)
,
\end{align*}
where $\textrm{D}_{\textrm{KL}}$ denotes the Kullback-Leibler
divergence 
\[\textrm{D}_{\textrm{KL}}( P | | Q) = \int{
\log\frac{P(x)}{Q(x)} \dif{P(x)}}.\]
\end{fact}
\noindent
In order to use this theorem, we have to bound the eigenvalues of
$H(x)H(y)^{-1}.$ For $x,y$ that are close in the local norm, this
follows since $H(x) \approx H(y).$
\begin{proofof}{Lemma~\ref{lem:gaussian-stat-dist}}
From the assumption, we have,
\[\frac{c^2}{n} \ge \norm{x-y}_{x}^{2} = \sum_{i \in [m]}
\frac{(a_{i}^{\top}(x-y))^{2}} {(a_{i}^{\top} x - b_{i})^{2}} \ge \max_{i \in [m]}
\frac{(a_{i}^{\top}(x-y))^{2}} {(a_{i}^{\top} x - b_{i})^{2}}.\]
Thus, for all $i \in [m],$ we have
\[ \left(1-\frac{c}{\sqrt{n}}\right) (a_{i}^{\top} x - b_{i}) \le
(a_{i}^{\top} y - b_{i})\le \left(1 + \frac{c}{\sqrt{n}}\right)
(a_{i}^{\top} x - b_{i}).\] 
By the definition of $H,$ we get,
\[ \left(1-\frac{c}{\sqrt{n}}\right)^{2} H(y)\preceq H(x) \preceq \left(1 +
  \frac{c}{\sqrt{n}}\right)^{2} H(y).\]
Thus, all eigenvalue $\lambda_{1},\ldots,\lambda_{n} > 0$ of
$H(x)H(y)^{-1},$ satisfy 
\[ \left(1-\frac{c}{\sqrt{n}}\right)^{2} \le \lambda_{i} \le \left(1 +
  \frac{c}{\sqrt{n}}\right)^{2} .\]
We can now bound the statistical distance between
$g_{x} = \calN\left(x, \frac{r^{2}}{n} H(x)^{-1} \right)$ and
$g_{y} = \calN\left(y, \frac{r^{2}}{n} H(y)^{-1} \right)$ by using
Pinsker's inequality~\cite[p. 44]{CsiszarK11}, which gives that
$\norm{g_{x} - g_{y}}_{1}^{2} \le 2\cdot 
\textrm{D}_{\textrm{KL}}(g_{y} | | g_{x}).$
Letting $\Sigma_{1}, \Sigma_{2}$ denote the covariance matrices of
$g_{1}, g_{2},$ we can write $\tr{\Sigma_{1}^{-1} \Sigma_{2}} =
\sum_{i=1}^{n} \lambda_{i},$ and $\log \frac{\det \Sigma_{1}}{\det
  \Sigma_{2}} = \log \frac{1}{\det \Sigma_{1}^{-1} \Sigma_{2}} =
\sum_{i=1}^{n} \log \frac{1}{\lambda_{i}}.$ 
\begin{align*}
& \norm{g_{x} - g_{y}}^{2}_{1} \le  \sum_{i=1}^{n} \left(
                               \lambda_{i} - 1 + \log \frac{1}{\lambda_{i}}
\right) +
  \frac{n}{r^{2}}\norm{x-y}_{x}^{2}  
\intertext{\hfill (Using Fact~\ref{fact:multigaussian-KL})}
&  \qquad \le \sum_{i=1}^{n} \left( \lambda_{i} +
  \frac{1}{\lambda_{i}} - 2\right) +
  \frac{n}{r^{2}}\norm{x-y}_{x}^{2} 
\intertext{\hfill $\left( \textrm{Using }\log
  \frac{1}{\lambda} \le \frac{1}{\lambda}-1 \right)$}
&  \qquad  \le n\cdot \max\left\{ \frac{(\nfrac{2c}{\sqrt{n} -
  \nfrac{c^2}{n}})^2}{(1 - \nfrac{c}{\sqrt{n}})^2} , \frac{(\nfrac{2c}{\sqrt{n} +
  \nfrac{c^2}{n}})^2}{(1 + \nfrac{c}{\sqrt{n}})^2} \right\} +
  \frac{n}{r^2}\cdot \frac{c^2}{n}.
\intertext{\hfill $\left(\textrm{Using the convexity of }\lambda +
  \frac{1}{\lambda}-2 \right)$}
&  \qquad \le n \cdot \frac{(\nfrac{2c}{\sqrt{n} -
  \nfrac{c^2}{n}})^2}{(1 - \nfrac{c}{\sqrt{n}})^2} +
  \frac{c^2}{r^2}\\
&  \qquad = c^2\cdot \frac{(2 - \nfrac{c}{\sqrt{n}})^2}{(1 - \nfrac{c}{\sqrt{n}})^2} +
  \frac{c^2}{r^2} \le \frac{25}{4}c^2 +  c^2 \le 9c^2,
\end{align*}
where the last line uses $c \le \nfrac{1}{3}, r \le 1$ and $n \ge 1.$
\end{proofof}

\section{The effect of the Metropolis filter} 
In this section, we prove Lemma~\ref{lem:metropolis-stat-dist} that
shows that for any $x \in K,$ the statistical distance between the
Gaussian distribution $g_{x}$ and the random walk distribution $p_{x},$
obtained by applying the Metropolis filter to $g_{x},$ is small. We
have,
\begin{align}
\label{eq:metropolis-stat-dist}
  \norm{p_x(z) - g_{x}(z)}_{1} = 1 - \av_{z \sim g_{x}} 
  \min\left\{1,\frac{g_z(x)}{g_x(z)}\right\}.
\end{align}
Given $\eps \in (0,\nfrac{1}{2}],$ we show that for an appropriate
choice of $r,$ the above statistical distance is bounded by $\eps.$

The ratio of $g_{z}$ and $g_{x}$ has two terms: one involving the
ratio of $\det H(x)$ and $\det H(z),$ and one involving the difference
in local norms $\norm{z-x}_{z}^{2} - \norm{z-x}^{2}_{x}.$
Proposition~\ref{lem:log-det} bounds the first by controlling the norm of
$\nabla \log \det H(x).$ Proposition~\ref{lem:local-norm-change} bounds the
second term by using concentration of Gaussian polynomials.
\begin{proofof}{of Lemma~\ref{lem:metropolis-stat-dist}}
We have,
\begin{align*}
& \frac{g_z(x)}{g_x(z)} = 
\exp\left( -\frac{n}{2r^2} \left(\norm{z-x}_{z}^2 - \norm{z-x}^{2}_{x}
  \right) \right. \\
& \qquad \qquad \qquad \qquad \left. + \frac{1}{2}\left( \log \det H(z) - \log \det H(x)\right)
\right).
\end{align*}
From Proposition~\ref{lem:log-det}, for
$r \le \frac{\eps}{4} (2\log\nfrac{4}{\eps})^{-\nfrac{1}{2}},$ we have
\[\pr[\log \det H(z) - \log \det H(x) \ge -\nfrac{\eps}{2}] \ge
1-\nfrac{\eps}{4}.\]
Also, from Proposition~\ref{lem:local-norm-change}, for
$r \le \frac{\eps}{100} (\log \nfrac{50}{\eps} )^{-\nfrac{3}{2}},$ we
have,
\[\pr\left[ \norm{z-x}^2_{z} - \norm{z-x}^{2}_{x} \le
  \frac{\eps}{2}\cdot\frac{r^2}{n} \right] \ge 1 - \nfrac{\eps}{4}.\]
Combining the two using a union bound, we get that except with
probability $\nfrac{\eps}{2},$ we have,
$\frac{g_z(x)}{g_x(z)} \ge e^{-\nfrac{\eps}{2}} \ge
1-\nfrac{\eps}{2}.$ Thus, 
\begin{align*}
  \av_{z \sim g_{x}} \min\left\{ 1,
  \frac{g_z(x)}{g_x(z)}\right\} 
  & \ge \left(1-\frac{\eps}{2}\right)
    \pr_{z \sim g_{x}} \left[\frac{g_z(x)}{g_x(z)} \ge
    1-\frac{\eps}{2}\right]
\\ & \ge \left(1-\frac{\eps}{2}\right)^2 \ge
    1-\eps.
\end{align*}
The claim now follows from~\eqref{eq:metropolis-stat-dist}.
\end{proofof}

\begin{proposition}
\label{lem:log-det}
Given $\eps \in (0,\nfrac{1}{2}],$ for
$r \le \frac{\eps}{\sqrt{2\log\nfrac{1}{\eps}}},$ and $z \sim g_{x}$
we have
\[\pr[\log \det H(z) - \log \det H(x) \ge -2\eps]
\ge 1-\eps.\]
 \end{proposition}
\begin{proofof}{of Proposition~\ref{lem:log-det}}
  Let $V(x) \defeq \frac{1}{2}\log \det H(x).$ From the work of Vaidya~\cite{Vaidya96}, we know that $V(x)$ is a convex
  function. Thus, $V(z) - V(x) \ge (z-x)^{\top}\nabla V(x).$ We know
  that
  $z = x + \frac{r}{\sqrt{n}} \left(H(x) \right)^{-\nfrac{1}{2}} g,$
  where $g \sim \calN(0,\Id{n}).$ Thus,
  \[V(z) - V(x) \ge \frac{r}{\sqrt{n}} g^{\top}\left(H(x)
  \right)^{-\nfrac{1}{2}} \nabla V(x).\]
  $g^{\top}\left(H(x) \right)^{-\nfrac{1}{2}} \nabla V(x)$ is a
  Gaussian with mean 0 and variance
  $\norm{\left(H(x) \right)^{-\nfrac{1}{2}} \nabla V(x)}^{2}_{2}.$
  From Lemma 4.3 in the work of Vaidya and Atkinson~\cite{VaidyaA93}, it follows that
\[\norm{\left(H(x) \right)^{-\nfrac{1}{2}} \nabla V(x)}^{2}_{2} \le
n.\] Using standard tail bounds, we get that for all $\lambda > 0,$
\[\pr[g^{\top}\left(H(x) \right)^{-\nfrac{1}{2}} \nabla V(x) \ge -\lambda
\sqrt{n}] \ge 1- \exp({-\nfrac{\lambda^2}{2}}).\]
Picking $\lambda = \sqrt{2\log \nfrac{1}{\eps}},$ and combining, we
get,
$\pr[V(z) - V(x) \ge -r\sqrt{2\log \nfrac{1}{\eps}}] \ge 1-\eps.$ For
$r \le \frac{\eps}{\sqrt{2\log\nfrac{1}{\eps}}},$ we have
$r\sqrt{2\log \nfrac{1}{\eps}} \le \eps,$ which gives the claim.
\end{proofof}
\begin{proposition}
\label{lem:local-norm-change}
Given $\eps \in (0,\nfrac{1}{2}],$ for
$r \le \frac{\eps}{20} (\log \nfrac{11}{\eps}
)^{-\frac{3}{2}},$ and $z \sim g_{x},$ we have,
\[\pr\left[ \norm{z-x}^2_{z} - \norm{z-x}^{2}_{x} \le
  2\eps\cdot\frac{r^2}{n} \right] \ge 1 - \eps.\]
\end{proposition}

\begin{proofof}{Proposition~\ref{lem:local-norm-change}}
  We have
  $z = x + \frac{r}{\sqrt{n}} \left(H(x)
  \right)^{-\frac{1}{2}} g,$
  where $g \sim \calN(0,\Id{n}).$ If we let
  $\hat{a}_i = \frac{1}{a_{i}^{\top}x - b_{i}}\left(H(x)
  \right)^{-\frac{1}{2}} a_{i},$
  we get
  ${a}_{i}^{\top}(z-x) = \frac{r}{\sqrt{n}}(a_{i}^{\top}x - b_{i})\cdot
  \hat{a}_{i}^{\top}g,$
  and $\sum_{i=1}^m \hat{a}_i \hat{a}_i^{\top} = \mathbbm{I}_{n}.$
\begin{align}
\label{eq:local-norm-change:terms}
 &  \norm{z-x}^2_{z} - \norm{z-x}^{2}_{x}  \nonumber
  \\
& = \sum_{i=1}^m (a_{i}^{\top}(z-x))^2 \left(
    \frac{1}{(a_{i}^\top z - b_{i})^2 } -  \frac{1}{(a_{i}^\top x -
    b_{i})^2 } \right) \nonumber\\
  & = \frac{r^2}{n}\sum_{i=1}^m (\hat{a}_{i}^{\top}g)^2 \left(
    \frac{1}{(1+\frac{r}{\sqrt{n}}\hat{a}_{i}^\top g)^2 } -1 \right)
\nonumber    \\
  & =  \frac{r^4}{n^2} \sum_{i=1}^m 
    (\hat{a}_{i}^{\top}g)^4   \left(\frac{2}{(1 + \frac{r}{\sqrt{n}}\hat{a}_{i}^\top
    g)}\right. + \left.\frac{1}{(1 + \frac{r}{\sqrt{n}}\hat{a}_{i}^\top g)^2}
    \right)
 \nonumber \\
& \qquad \qquad
- \frac{2r^3}{n^{\nfrac{3}{2}}}\sum_{i=1}^m
    (\hat{a}_{i}^{\top}g)^3. 
\end{align}
We now use concentration of Gaussian polynomials  (see Theorem~\ref{thm:concentration}) to bound the
two terms above. Let
$P_{1}(g) \defeq \sum_{i=1}^m (\hat{a}_{i}^{\top}g)^3.$ From
Fact~\ref{lem:homogeneous-poly}, we know 
$\av_{g} P_{1}(g)^{2} \le 15n.$ Thus, using
Theorem~\ref{thm:concentration}, we know that for any
$\lambda_1 \ge (\sqrt{2e})^{3},$
\[\pr_{g}\left[|P_{1}(g)| \ge \lambda_1 \sqrt{15n}\right] \le
\exp\left(-\frac{3}{2e} \lambda_{1}^{\nfrac{2}{3}}\right).\]
Picking
$\lambda_{1} = \left(\max\left\{2e,\frac{2e}{3} \log
    \frac{2}{\eps}\right\}\right)^{\nfrac{3}{2}},$
and $r \le \frac{\eps}{2\sqrt{15}\lambda_{1}},$ we obtain,
$\pr\left[|P_{1}(g)| \ge \frac{\eps}{2r}\sqrt{n}\right] \le
\frac{\eps}{2}.$ Thus, with probability at least $1-\frac{\eps}{2},$
\begin{align}
\label{eq:local-norm-change:term1}
-\frac{2r^3}{n^{\nfrac{3}{2}}}\sum_{i=1}^m
  (\hat{a}_{i}^{\top}g)^3 \le \frac{2r^3}{n^{\nfrac{3}{2}}} \cdot
  \frac{\eps}{2r}\sqrt{n} = \eps\cdot\frac{r^2}{n}.
\end{align}

\noindent
Now, we let $P_{2}(g) \defeq \sum_{i=1}^m (\hat{a}_{i}^{\top}g)^4.$
Again, from Fact~\ref{lem:homogeneous-poly}, we know that
$\av_{g} P_{2}(g)^{2} \le 105n^2,$ and applying
Theorem~\ref{thm:concentration}, we obtain that for 
$\lambda_{2} = \left(\max\left\{2e,\frac{2e}{4} \log
  \frac{2}{\eps}\right\}\right)^{2}$
and $r \le \frac{\sqrt{\eps}}{\sqrt{8\lambda_{2} \sqrt{105}}},$ we obtain,
$\pr\left[|P_{2}(g)| \ge \frac{\eps}{8r^2}n\right] \le
\frac{\eps}{2}.$ Thus, with probability at least $1-\frac{\eps}{2}$.
\begin{align*}
& \frac{r^4}{n^2} \sum_{i=1}^m 
(\hat{a}_{i}^{\top}g)^4 \le \frac{r^4}{n^2} \cdot \frac{\eps}{8r^2}n =
\frac{\eps}{8}\cdot\frac{r^2}{n}. 
\end{align*}

Note that this also implies that for all $i,$
$\frac{r}{\sqrt{n}} |\hat{a}_{i}^{\top}g| \le \left(\frac{\eps
    r^{2}}{8n}\right)^{\nfrac{1}{4}} \le \frac{1}{2},$
where the last inequality holds for all $r \le 1.$ Thus, with
probability at least $1-\frac{\eps}{2},$ we have
\begin{align*}
& \frac{r^4}{n^2} \sum_{i=1}^m 
(\hat{a}_{i}^{\top}g)^4 \left(\frac{2}{(1 +
    \frac{r}{\sqrt{n}}\hat{a}_{i}^\top g)} + \frac{1}{(1 +
    \frac{r}{\sqrt{n}}\hat{a}_{i}^\top g)^2} \right) \\
& \qquad \qquad \le 8 \frac{r^4}{n^2} \sum_{i=1}^m 
    (\hat{a}_{i}^{\top}g)^4 \le \eps\cdot\frac{r^2}{n}.
\end{align*}
Combining this with Equations~\eqref{eq:local-norm-change:terms} and
\eqref{eq:local-norm-change:term1}, and applying a union bound, we get
that with probability at least $1-\eps$
\[  \norm{z-x}^2_{z} - \norm{z-x}^{2}_{x} \le 2\eps\cdot\frac{r^2}{n} 
.\]
\noindent
Finally, we verify that for $\eps \in (0,\nfrac{1}{2}],$ any $r
\le \frac{\eps}{20} (\log \nfrac{11}{\eps} )^{-\frac{3}{2}}$
satisfies the conditions $$r \le \min \left\{1,
\frac{\eps}{2\sqrt{15}\lambda_{1}},
\frac{\sqrt{\eps}}{\sqrt{8\lambda_{2} \sqrt{105}}} \right\}.$$
\end{proofof}

\begin{theorem}
\textbf{(see Janson~\cite[Thm 6.7]{Janson97})}
\label{thm:concentration} 
  Let $P(g)$ be a degree $q$ polynomial, where $g \in \rea^n$ such
  that $g\sim \calN(0,\Id{n}).$ Then, for any $t \ge \sqrt{2e}^{q},$
  we have,
\[\pr_{g}\left[|P(g)| \ge t\left(\av P(g)^{2}\right)^{\nfrac{1}{2}} \right] \le \exp\left(-\frac{q}{2e}t^{\nfrac{2}{q}}\right).\]
\end{theorem}

\section*{Acknowledgements}
The work of the first author was supported by a Simons
      Investigator Award to Daniel Spielman.

\appendix
\section{Relating the local metric to the Hilbert metric}
\begin{lemma}[\cite{KannanN2012}]
\label{lem:hilbert-local-metric}
For any $x,y \in K,$ we have
$\sigma(x,y) \ge \frac{1}{\sqrt{m}} \norm{x-y}_{x}.$
\end{lemma}
\begin{proofof}{Lemma~\ref{lem:hilbert-local-metric}}
Let $p,x,y,q$ be the points in order on the chord of $K$ that passes
through $x,y$ with $p,q$ being the end-points of the chord.
Thus,
\begin{align*}
\sigma(x,y) =  \frac{|x-y||p-q|}{|p-x||q-y|} & \ge \max\left\{ \frac{|x-y|}{|p-x|}, \frac{|x-y|}{|x-q|}\right\} \\
& = \max_{i \in [m]} \frac{|a_{i}^{\top}(x-y)|}{(a_{i}^{\top}x -
  b_{i})}  \\
& \ge \frac{1}{\sqrt{m}} \left(  \sum_{i \in [m]}
\frac{(a_{i}^{\top}(x-y))^{2}} {(a_{i}^{\top} x - b_{i})^{2}}
  \right)^{\nfrac{1}{2}} \\
& = \frac{1}{\sqrt{m}} \norm{x-y}_{x}.
\end{align*}
\end{proofof}

\section{Moments of Gaussian Polynomials}
\begin{fact}
\label{lem:homogeneous-poly}
  Suppose $g\in \rea^n$ is distributed according to $\calN(0,\Id{n}),$
  and $\sum_{i=1}^m b_i b_i^\top = \mathbbm{I}_{n}.$ Then, we
  have,
\[\av \left( \sum_{i=1}^m (b_{i}^\top g)^3 \right)^2 \le 15n, 
\  \textrm{ and } \ 
\av \left( \sum_{i=1}^m (b_{i}^\top g)^4 \right)^2 \le 105n^2.\]
\end{fact}
\begin{proofof}{of Fact~\ref{lem:homogeneous-poly}}
We first consider the first part of the fact. From
Fact~\ref{lem:monomial-av}, we know that for all $i,j,$ 
\[ \av_g (b_{i}^\top g)^3 (b_{j}^\top g)^3 = 9 \norm{b_{i}}^2
\norm{b_{j}}^2 (b_{i}^{\top} b_{j}) + 6 (b_{i}^{\top} b_{j})^3.\]
Summing over all $i,j,$ we get,
\begin{align}
&  \av \left( \sum_{i=1}^m (b_{i}^\top g)^3 \right)^2 
  = \sum_{i,j=1}^m
    \av (b_{i}^\top g)^3 (b_{j}^\top g)^3 \nonumber \\
& \qquad     =   9 \sum_{i,j=1}^m  \norm{b_{i}}_{2}^2
    \norm{b_{j}}_{2}^2 (b_{i}^{\top} b_{j}) + 6 \sum_{i,j=1}^m
    (b_{i}^{\top} b_{j})^3.
\label{eq:homogeneous-poly:terms}
\end{align}
\noindent This equality can also be derived using Isserlis' theorem (\cite{Isserlis1918}).
If we let $B$ be the $m \times n$
matrix with its $i^\textrm{th}$ row being $b^{\top}_{i},$ and $w \in \rea^m$ be such
that $w_{i} = \norm{b_{i}}_{2}^{2},$ we can simplify the first term in
the above sum as follows. 
\begin{align*}
\sum_{i,j=1}^m  \norm{b_{i}}_{2}^2
    \norm{b_{j}}_{2}^2 (b_{i}^{\top} b_{j}) = \norm{\sum_{i=1}^m \norm{b_{i}}^2 b_{i}}_{2}^2 = \norm{B^{\top}w}_{2}^{2}.
\end{align*}
Using $\sum_{i=1}^m b_i b_i^\top = \Id{n},$ we get
$B^{\top} B = \Id{n}.$ Thus, the $m \times m$ matrix
$\Pi \defeq BB^{\top}$ satisfies $\Pi^2 = \Pi.$ Since $\Pi$ is also
symmetric, it is an orthogonal projection. Thus, we have
$\norm{\Pi w}_{2} \le \norm{w}_{2}.$ We obtain,
\begin{align*}
  \norm{B^{\top}w}_{2}^{2} =
  w^{\top} BB^{\top} w & = w^\top\Pi w \\
& = w^\top \Pi^{2} w  \\
& = \norm{\Pi
  w}_{2}^2 \le \norm{w}_{2}^{2} = \sum_{i=1}^m \norm{b_{i}}^4.
\end{align*}
Since $\sum_{i=1}^m b_i b_i^\top = \mathbbm{I}_{n},$ we get that for
all $i,$ $\norm{b_{i}} \le 1.$ Moreover, taking trace, we obtain
$\sum_{i=1}^m \norm{b_{i}}^{2} = n.$ Thus,
\[\sum_{i=1}^m \norm{b_{i}}^4 \le \sum_{i=1}^m \norm{b_{i}}^2 = n.\]
Thus, we can bound the first term in
Equation~\eqref{eq:homogeneous-poly:terms} by $n.$ 

For the second term in Equation~\eqref{eq:homogeneous-poly:terms},
using $\norm{b_{i}}_{2} \le 1$ for all $i,$ and Cauchy-Schwarz, we get
$|b_{i}^\top b_{j}| \le 1$ for all $i,j.$ Using
$\sum_{i=1}^m b_i b_i^\top = \Id{n},$ we also know that for all $j,$
$\sum_{i=1}^m (b_{i}^\top b_{j})^{2} = \norm{b_{j}}_{2}^{2}.$ Thus, we
get,
\[\sum_{i,j=1}^m (b_{i}^{\top} b_{j})^3 \le \sum_{i,j=1}^m
(b_{i}^{\top} b_{j})^2 = \sum_{j=1}^{m} \norm{b_{j}}_{2}^{2} = n.\]
Combining the bounds for the two terms in
Equation~\eqref{eq:homogeneous-poly:terms}, we get the first part of
the fact.

\medskip
For the second part of the fact, we use Cauchy-Schwarz inequality,
\begin{align*}
  \av \left( \sum_{i=1}^m (b_{i}^\top g)^4 \right)^2 
&  = \sum_{i,j=1}^m
  \av (b_{i}^\top g)^4 (b_{j}^\top g)^4 \\
&   \le \sum_{i,j=1}^m \left(\av (b_{i}^\top g)^8\right)^{\nfrac{1}{2}}
  \left(\av (b_{j}^\top g)^8\right)^{\nfrac{1}{2}}.
\end{align*}
We have that $b_{i}^\top g$ is distributed as a Gaussian with mean 0
and variance $\norm{b_{i}}_{2}^{2}.$ Thus,
$\av (b_{i}^\top g)^8 = 105 \norm{b_{i}}_{2}^8.$ Hence, we get,
\begin{align*}
\av \left( \sum_{i=1}^m (b_{i}^\top g)^4 \right)^2 & \le 105
\sum_{i,j=1}^m \norm{b_{i}}_{2}^4 \norm{b_{j}}_{2}^4 \\
& = 105 \left(
  \sum_{i=1}^m \norm{b_{i}}^4 \right)^{2} \le 105 n^2,
\end{align*}
proving the second part of the fact.
\end{proofof}
{\small \textbf{Note:}
The bounds given by the above fact are tight for the case where
$b_{i}$ form an orthonormal basis.
}

\pagebreak
\begin{fact}[Isserlis~\cite{Isserlis1918}]
\label{lem:monomial-av}
Suppose $g\in \rea^n$ is distributed according to $\calN(0,\Id{n}),$
and $b_1,b_2$ are any two vectors in $\rea^n,$
\begin{align*}
  \av_g  (b_{1}^\top g)^3 (b_{2}^\top g)^3 
  & = 9 \norm{b_{1}}^2
    \norm{b_{2}}^2 (b_{1}^{\top} b_{2}) + 6 (b_{1}^{\top} b_{2})^3.
\end{align*}
\end{fact}
\begin{proofof}{of Fact~\ref{lem:monomial-av}}
{We define $\hat{b}_{i} \defeq \frac{1}{\norm{b_{i}}} \cdot b_{i}$ to be the
corresponding unit vectors.}
 Thus,
\begin{align}
\label{eq:monomial-av:scaling}
\av  (b_{1}^\top g)^3 (b_{2}^\top g)^3 = \norm{b_{1}}^3
\norm{b_{2}}^3\av (\hat{b}_{1}^\top g)^3 (\hat{b}_{2}^\top g)^3.
\end{align}
We let $e_{1}, \ldots, e_{n}$ denote the standard basis vectors for
$\rea^n$, \emph{i.e.}, $e_{i}$ is 1 in the $i^\textrm{th}$ coordinate
and 0 elsewhere. Since the distribution of $g$ is rotationally
symmetric, we can assume that $\hat{b}_{1} = e_{1},$ and $\hat{b}_{2}
= \cos \theta \cdot e_{1} + \sin \theta \cdot e_{2},$ where $\theta$ is such that
$\cos \theta = \hat{b}^{\top}_{1} \hat{b}_{2}.$ Thus,
\begin{align*}
& \av (\hat{b}_{1}^\top g)^3 (\hat{b}_{2}^\top g)^3  = \av g_{1}^3 \left(
  \cos \theta \cdot g_{1} + \sin \theta \cdot g_{2} \right)^3 \\
& \qquad \qquad   = \cos^3 \theta \av g_{1}^6 + 0 + 3 \cos \theta \sin^2 \theta \av
  g_{1}^4 \av g_{2}^2 + 0 \\
& \qquad \qquad  = 15 \cos^3 \theta + 9 \cos \theta \sin^2 \theta =  9
  \cos \theta  + 6 \cos^3 \theta \\
& \qquad \qquad  = 9 \left( \hat{b}^{\top}_{1} \hat{b}_{2} \right) + 6 \left( \hat{b}^{\top}_{1} \hat{b}_{2} \right)^3.
\end{align*}
Combining with Equation~\eqref{eq:monomial-av:scaling}, we obtain the fact.
\end{proofof}

\end{document}